\documentclass[reprint,amsmath,amssymb,aps]{revtex4}
\usepackage{graphicx}
\usepackage{pdfpages}
\usepackage{epsfig,natbib,color}
\usepackage{lscape}
\usepackage{graphicx}
\usepackage{epsfig}
\usepackage{natbib}
\usepackage{textcomp}
\usepackage{gensymb}
\usepackage{multirow}
\usepackage{hyperref}
\renewcommand\apj{{Astrophys. J.}}
\newcommand\apjl{{Astrophys. J. Lett.}}     
\newcommand\apjs{{Astrophys. J. Suppl. Ser.}}
\newcommand\aap{{Astron. Astrophys.}}
\newcommand\mnras{{Mon. Not. Roy. Astron. Soc.}}
\newcommand\solphys{{Sol. Phys.}}
\newcommand\ssr{{Space Sci. Rev.}}
\renewcommand\nat{{Nature}}

\newcommand{\unit}[1]{\ensuremath{\,\mathrm{#1}}}
\def\kms{$\rm{km~s}^{-1}$}
\begin{document}
2020, Nature Astronomy, 4, 994  \hspace{2cm}  
\url{https://www.nature.com/articles/s41550-020-1094-3}
\\
\\
\title{Simulations of solar filament fine structures and their counterstreaming flows}
\author{Y. H. Zhou$^{1, 2, 3}$,  P. F. Chen$^{1, 3}$\footnote{Corresponding author, chenpf@nju.edu.cn}, J. Hong$^{1, 3}$  \& C. Fang$^{1, 3}$}
\affiliation{$^{1}$School of Astronomy and Space Science, Nanjing University, Nanjing 210023, China\\}
\affiliation{$^2$Centre for mathematical Plasma Astrophysics, Department of Mathematics, KU Leuven, Celestijnenlaan 200B, 3001 Leuven, Belgium\\}
\affiliation{$^3$Key Laboratory of Modern Astronomy and Astrophysics (Nanjing University), Ministry of Education, China}

\begin{abstract}
Solar filaments, also called solar prominences when appearing above the solar limb, are cold, dense materials suspended in the hot tenuous solar corona, consisting of numerous long, fibril-like threads. These threads are the key to disclosing the physics
of solar filaments. Similar structures also exist in galaxy clusters. Besides their mysterious formation, filament threads are observed to move with alternating directions, which are called counterstreaming flows. However, the origin of these flows has not been clarified yet. Here we report that turbulent heating at the solar surface is the key, which randomly evaporates materials from the solar surface to the corona, naturally reproducing the formation and counterstreamings of the sparse threads in
the solar corona. We further suggest that while the cold H$\alpha$ counterstreamings are mainly due to longitudinal oscillations of the filament threads, there are million-kelvin counterstreamings in the corona between threads, which are alternating unidirectional flows.
\end{abstract}
\maketitle
\section*{Introduction}
Solar filaments, also called prominences, are elongated cold dense plasma structures embedded in the hot tenuous solar corona \citep{mack2010, vial2015}. Once erupting, they are accompanied by giant arcades, solar flares and/or coronal mass ejections (CMEs) \citep{chen2011, schm2013}. It has been proposed that the filament plasma originates from the solar chromosphere, where the cold materials can be injected directly to the corona \citep{wang1999, wang2019}, or are heated to several million degrees and evaporate into the corona, and then thermal non-equilibrium leads to catastrophic cooling and plasma condensation \citep{anti1991, liu2012}. It might also emerge from the sub-surface to the corona \citep{lite1997} or result from magnetic reconnection in the corona \citep{kane2017, li2019}.

While these processes can explain how materials with a mass up to $10^{14}$--$10^{15}$ g \citep[ref. ][]{mack2010} accumulate locally in the corona, there is no consensus on the formation of the fibril structures of a filament. These fibrils are called threads \citep{dunn1961, simo1986, hein2007}. High-resolution observations in H$\alpha$ 6562.8 \AA\ revealed that the thread has a typical width of 100--300 km and a length of several to 20 Mm \citep[refs.][]{engv1998, lin2005}.

One possibility is that the hosting magnetic field is finely structured. For example, Hood et al. \citep{hood1992} extended the traditional magnetic configuration with one dip to a model with a series of dips. While there might be 2 or 3 dips along a single magnetic field line \citep{zhou2017}, it has not been demonstrated that there exists a series of dips along one field line \citep{vial2015}. Later, filament threads were proposed to be formed by plasma instabilities \citep{prie1991, vand1993, hill2016}. While some of them can explain the typical width of filament threads in observations, these models are mainly based on linear analysis, and it is not clear whether the threads can sustain for long. Furthermore, recent 3-dimensional (3D) magnetohydrodynamic (MHD) simulations suggested that filament threads might be an apparent structure, which is strongly mis-aligned with the local magnetic field \citep{xia2016}. This posed a strong challenge to the traditional understanding on filament threads \citep{hein2006, mart2008}. Further simulations and observations \citep{schm2017} are required to examine the relation between filament threads and the local magnetic field.

Another intriguing feature of solar filaments is the so-called counterstreamings, i.e., plasma moves with a typical velocity of $\sim$20 \kms\ in opposite directions between neighbouring threads \citep{schm1991, zirk1998, lin2003, wang2018}. They may be due to independent longitudinal oscillations of different threads \citep{lin2003} and/or alternating unidirectional flows \citep{chen2014}, both scenarios were confirmed by later observations \citep{ahn2010, zou2016}. Unequal heating at the two conjugate footpoints of a magnetic field line results in the oscillation of filament threads, leading to counterstreamings, as confirmed by either 1-dimensional (1D) simulations \citep{anti1991} or pseudo 3D rendering of an assemble of 1D simulations \citep{luna2012a}. The counterstreamings were also detected in extreme ultraviolet (EUV) spectral lines, which correspond to plasma with the temperature around $10^5-10^6$ degrees \citep[refs.][]{wang1999, kuce2003, kuce2014, alex2013, dier2018}. A puzzling characteristic of the EUV counterstreamings is that their velocity can be up to 100 \kms \citep[ref.][]{alex2013}. How the slow H$\alpha$ cold flows and the fast EUV hot flows fit into a self-consistent picture deserves further exploration.

With the purpose to interpret the origin and dynamics of filament threads, we plan to investigate the response of the solar corona to turbulent heating on the solar surface.

\section*{Thread formation}
The magnetic configuration is illustrated in Fig. \ref{fig1}. After the localised random heating is imposed at the footpoints of the magnetic arcade for about 44 minutes, which drives chromospheric evaporation continually, the first filament thread appears near $y=-1.6$ Mm, as indicated by the temperature distribution in Fig.~\ref{fig2}a. Here we define an area as filament material if the radiative cooling makes the plasma temperature fall below 14,000 K \citep[ref.][]{hein2015}. The onset time of the condensation in our simulation is earlier than those in the previous works, which are usually longer than 2 \unit{hrs} \citep[ref.][]{xia2011}. One possible reason is that once condensation happens, the gas pressure decreases, and the flux tube is compressed further, which enhances condensation, forming a positive feedback.

As time goes on, more filament materials are formed, and the total mass of filament threads grows. Fig.~\ref{fig2}b shows the time evolution of the ratio of the area of the filament threads $A_{fila}$ to the total area of the initial coronal plasma $A_{total}$ in our simulation. It is seen that immediately after the first thread forms, $A_{fila}$ grows quickly and then keeps around 10\% for $\sim$200 minutes. During $t = 250$--300 minutes, the thread coverage dramatically increases again, and then fluctuates. Over all, the coverage of the filament threads, or the filling factor, is basically in the range of 10--15\% after $t = 100$ minutes in our simulation. We also check the averaged plasma density of the threads. As shown by the blue dashed line in Fig.~\ref{fig2}c, the plasma density grows gradually to be over 100 times the background coronal density, which is consistent with observations. We also see that the filament reaches a quasi-stable state after about $t = 100$ minutes according to the above analysis.

Figure~\ref{fig3}a shows the temperature distribution at $t =259$ minutes near the centre of our simulation box. Note that the temperature is shown in a logarithmic scale in order to see the thread structures more clearly. In this panel, many cold thread-like structures can be identified. We cut a slice along the black dashed line at the centre of this panel to see the distribution of the plasma temperature and density more clearly. The results are shown in Fig.~\ref{fig3}b, where the red solid line corresponds to the temperature distribution, and the blue dashed line represents the plasma number density. From this panel, we can see more clearly that the flux sheet becomes fragmented automatically due to our random heating. In our simulation, the typical width of the thread structure is $\sim$100 km, which matches the observations very well. The length of these threads, from several to about 20 or 30 Mm, is also comparable to observations. If the dip along the magnetic field line is deeper, shorter threads are expected.

It should be pointed out that, in observations, what we actually see is neither temperature nor density, but emission. To make the results comparable to observations, an H$\alpha$ image at $t=259$ minutes is synthesized approximately and is shown in Fig.~\ref{fig3}c, where a background emission is included (see Methods for details). Despite the approximation of calculation, the H$\alpha$ intensity map is much cleaner than the temperature distribution and individual threads can be clearly distinguished.

The synthetic EUV 193~\AA\ image at the same time is also calculated and is shown in Fig.~\ref{fig3}d. To conduct the calculation, the radiative transfer equation is solved approximately (see Methods for details). The whole panel looks faint compared to the quiet corona mainly because the temperature is lower than that of the quiet region. The bright part is usually along a magnetic field line where the heating is strong, or some areas enveloping the threads, i.e., the so called prominence--corona transition region. In this panel, we can also identify the thread structures, though it seems harder to distinguish one thread from the other than in panel c. This is consistent with the observational fact that EUV filaments are usually observed to be wider and fuzzier than H$\alpha$ filaments and the H$\alpha$ channel is easier to see thread structures than EUV lines, as indicated by Aulanier \& Schmieder \citep{aula2002}. It also implies that some dark threads in EUV wavebands, e.g., 193 \AA, do not correspond to H$\alpha$ threads. They are merely cool plasmas, e.g., 0.2 MK as indicated by Fig. \ref{fig3}b.

By combining the three images in Figs. \ref{fig1}--\ref{fig3}, we can see that although a filament channel might occupy a significant volume in the corona, the cold material, represented by the filament threads, actually occupies only a small portion, about 10--15\%, which is clearly indicated by the H$\alpha$ image. For EUV channels, a filament occupies a bigger volume since the plasma in a wider region falls into the temperatures with weaker EUV emission. Some cool but not so cold regions could be observed to be parts of the EUV filament. Although the volume coverage of filament threads is consistent with observations, the filament threads in our synthesized H$\alpha$ image (Fig. \ref{fig3}c) seem more sparsely distributed in space compared to real H$\alpha$ observations. This is presumably due to the fact that we are simulating one layer, whereas real H$\alpha$ observations correspond to the superposition of many layers along the line-of-sight. To make the synthesized H$\alpha$ image look more realistic, an ideal way is to stack up tens of different simulations representing different layers. But considering that the heating is randomly distributed both spatially and temporally, we take a more convenient way here that is just to stack up the snapshots from different times in our simulation. The resulting synthesized H$\alpha$ map is displayed in Fig. \ref{fig4}. Note that we have degraded the spatial resolution to 0.3 arcsec. The filament threads resemble the H$\alpha$ observations much better. Similar to Gunar et al. \citep{guna2018}, the synthesized H$\alpha$ intensity is not uniform along a magnetic dip. The difference between their results and ours is that the individual threads in our paper are formed naturally.

Another finding from these synthesized images is that all the formed filament threads are almost oriented along the $x$-direction, manifested as elongated fibrils. In our simulation, the initial magnetic field is uniform, oriented in the $x-$axis. During evolution, the magnetic field changes by up to only 20\% due to the low gas to magnetic pressure ratio. That is to say, the formed filament threads are mainly aligned with the local magnetic field. Of course, scrutinising Fig. \ref{fig3}c carefully, we can find that the filament threads are not exactly along the local magnetic field, i.e., some threads might be apparent structures, where different segments are frozen in immediately neighbouring field lines. After checking the numerical results, it is found that some filament threads are misaligned with the local magnetic field by $\sim$2 degrees. That is also why two thin threads merge into one unity at several locations in Fig. \ref{fig3}c. However, during the whole evolution process, there is no any thread that is misaligned with the local magnetic field by more than several degrees. It means that filament threads can be considered to trace the local magnetic field in practice.

\section*{Dynamics}

Although Fig. \ref{fig2}b indicates that the filament thread coverage is rather stable during $t=$100--170 minutes as mentioned above, the system is still extremely dynamic all the time. Figure \ref{fig5}a depicts the horizontal velocity ($v_x$) distribution in the central part of our simulation domain, where red means that the velocity is towards the right and blue means that the velocity is towards the left. It is seen that the filament channel is full of dynamics, with the plasma velocity up to more than 80 \kms. It is found that the velocities of the cold filament plasma and the hot coronal plasma are very different. In order to see the difference, we plot the time evolution of the spatially averaged rightward velocity of the hot coronal plasma as the red line in Fig. \ref{fig5}b, whereas the counterpart of the cold filament material as the blue line. It is revealed that the averaged velocity of the coronal plasma can reach $\sim$70--80 \kms, but the averaged velocity of the filament threads is less than 30 \kms\ throughout the evolution.

A remarkable feature in Fig. \ref{fig5}a is the plasma motions with alternating red- and blue-shifts, which are the typical counterstreamings, i.e., our simulation naturally reproduces the observed counterstreamings. In order to see how the filament threads move in the counterstreaming flows, we select a slice at $y=-1.1$ Mm as indicated by the dotted-dashed line, and the time--distance diagram of the H$\alpha$ intensity is plotted in Fig. \ref{fig5}c. It is seen that once formed, the H$\alpha$ filament thread begins to oscillate along the magnetic field, with a period of $\sim$60 minutes. A magenta line is marked near the oscillating filament thread with a slope of 12 \kms, indicating that the velocity amplitude of the filament thread oscillation is around 12 \kms. It is noticed that the filament thread oscillation is ceaseless and the oscillation is not centred at an equilibrium position, both probably due to the random heating at the two conjugate footpoints of the magnetic field line. The oscillation amplitude does not change monotonically with time, though it is the largest at the very beginning when the filament thread is formed.

The corresponding time--distance diagram of the synthesized 193 \AA\ intensity is plotted in Fig. \ref{fig5}d. As expected, the H$\alpha$ filament thread as revealed in Fig. \ref{fig5}c is manifested as dark features in the 193 \AA\ diagram in Fig. \ref{fig5}d. However, the 193 \AA\ diagram shows additional dark features due to the plasma temperature out of the emission range of the 193 \AA\ band. In addition to the dark features, 193 \AA\ brightening is seen from time to time in Fig. \ref{fig5}d as well. As marked by the oblique magenta line, the velocity of the moving 193 \AA\ brightening is around 70 \kms. The 193 \AA\ bright flows can be along the $x$-direction, e.g., around $t=$190 minutes, or along the negative $x$-direction, e.g., around $t=$150 minutes. We check slices at other locations, and find that the 193 \AA\ flows with alternating directions exist along other slices as well, even though there are no H$\alpha$ cold threads present. This means that EUV counterstreamings are present in our numerical results as well. They result from weaker chromospheric evaporation that is not dense enough to experience catastrophic cooling in the corona.

\section*{Discussion}

One purpose of our simulations is to reproduce the fine structure of filaments. Considering that the solar lower atmosphere is full of turbulent convection and ephemeral or moving magnetic features, we point out that the localised chromospheric heating should be turbulent, looking random apparently. With the turbulent heating imposed to the footpoints of an arcade of magnetic field lines, in this paper we reproduced the thermal structures in a filament channel. The length of these H$\alpha$ threads is up to about 20--30 Mm, and their averaged width is around 100 km. Moreover, the volume coverage of the threads in the filament channel is about 10--15\%. All these characteristics are consistent with observations \citep{engv1998, lin2005, alex2013}. As the heating is doubled, the size of the threads does not change much, but the volume coverage is doubled. As implied by Fig. \ref{fig5}c, the length of the filament thread grows with time initially, and then saturates. The length is determined by the morphology of the magnetic dip, influenced by mass drainage due to unidirectional flows. The origin of the $\sim$100 km width of the threads is still a puzzle worth exploring further. Here, we tentatively propose our idea, i.e., when a thread with a finite width is formed, the internal gas pressure is reduced, and the internal magnetic field is compressed. As a result, the external magnetic field expands. The expansion of the neighbouring flux tubes helps hinder the external plasma from further cooling. It is noted that filaments in real observations exhibit other fine structures, e.g., vertical threads \citep{guna2018}, which require further investigations.

Solar filaments have a lifetime of several days to weeks. During their lifetime, counterstreamings can be observed by H$\alpha$ spectroscopy, even when the filament is in a globally static state \citep{schm1991}. It is generally believed that the H$\alpha$ filament counterstreamings are owing to filament thread longitudinal oscillations along the magnetic field \citep{lin2003}, which was reinforced by later observations \citep{ahn2010} and numerical simulations \citep{luna2012a}. Our Fig. \ref{fig5}a clearly shows that counterstreamings are present in our simulations due to random heating localised at the footpoints, as also suggested by Kucera et al. \citep{kuce2014}. While Fig. \ref{fig5}c further indicates that the H$\alpha$ counterstreamings are mainly due to filament thread longitudinal oscillations, with a typical velocity about 10--25 \kms, careful examination of the numerical results indicates that there exist alternating unidirectional flows with a velocity of 1--2 km s$^{-1}$ inside the cold threads as well, which leads to the drainage of some threads. Such a result confirms the proposal of Chen et al. \citep{chen2014} that H$\alpha$ counterstreamings are due to the combination of filament thread longitudinal oscillations and alternating unidirectional flows.

Our numerical simulation further indicates that there exist EUV counterstreamings with velocities up to 100 \kms, as revealed by the bright and dark oblique ridges in Fig. \ref{fig5}d. The bright ridges in the synthesized 193 \AA\ image correspond to the hot plasma flows, e.g., 1 million degrees, and the dark 193 \AA\ ridges correspond to cooler (but still warm) plasma flows, e.g., 0.1 million degrees, as indicated by Fig. \ref{fig3}b. These results are consistent with the multi-temperature flows in observations \citep{alex2013}. Besides the EUV counterstreamings along the magnetic field lines with H$\alpha$ threads, EUV counterstreamings are also present in the coronal loops without H$\alpha$ threads, which we call interthread loops. For example, in the space between adjacent threads in Fig. \ref{fig3}c, e.g., in the range of $1.5<y<2$ Mm or $-0.7<y<0.2$ Mm, we can always find alternating red- and blue-shifted flows in the coronal loops in Fig. \ref{fig3}a, with velocities $\sim$80 \kms. Recent observations indeed showed  high-speed (over 70--80 \kms) EUV flows with oppositely directed velocities in neighbouring EUV threads \citep{kuce2003, alex2013}, which are reproduced in our numerical simulations. It means that the turbulent heating localised at the solar surface not only drives the formation of sparse H$\alpha$ threads and ensuing oscillations, forming H$\alpha$ counterstreamings, but also drives EUV counterstreamings in the interthread corona.

To summarise, with heat conduction and optically-thin radiative cooling considered in 2D MHD simulations, we reproduced for the first time the formation of filament threads, with the length, width, and filling factor all comparable to observations. We confirm that filament threads trace the local magnetic field extremely well, with the deviation of alignment less than several degrees. The simulations also reproduced counterstreamings both in H$\alpha$ and EUV wavebands. Besides the EUV counterstreamings in the coronal part of the flux tubes associated with H$\alpha$ cold threads, we propose that there exist EUV counterstreamings in the interthread corona as well.


\acknowledgements
The authors thank P. Heinzel, R. Keppens, T. Yokoyama, K. E. Yang, and E. S. Liang for valuable suggestions, partly during the ISSI-BJ team meeting. P.F.C was supported by the Chinese foundations (NSFC 11533005 and 11961131002) and Jiangsu 333 Project (BRA2017359). Y.H.Z is supported by the Belgian FWO-NSFC project G0E9619N and the Program A for Outstanding PhD candidates in Nanjing University. The simulations were done on the computers in the High Performance Computing Centre of Nanjing University.

\section*{Author contributions}

Y.H.Z. did the simulations and wrote the first draft. P.F.C. initiated the study, supervised the project and led the discussions. J.H. contributed to the calculation of the synthesized EUV image. C.F. contributed to discussions on the H$\alpha$ calculation.

\section*{Additional information}

Correspondence and requests for information should be addressed to P.F.C. (chenpf@nju.edu.cn).

\textbf{Competing Financial Interests:} The authors declare no competing financial interests.

\clearpage
\begin{figure}
\centering
      \includegraphics[width=0.60\textwidth]{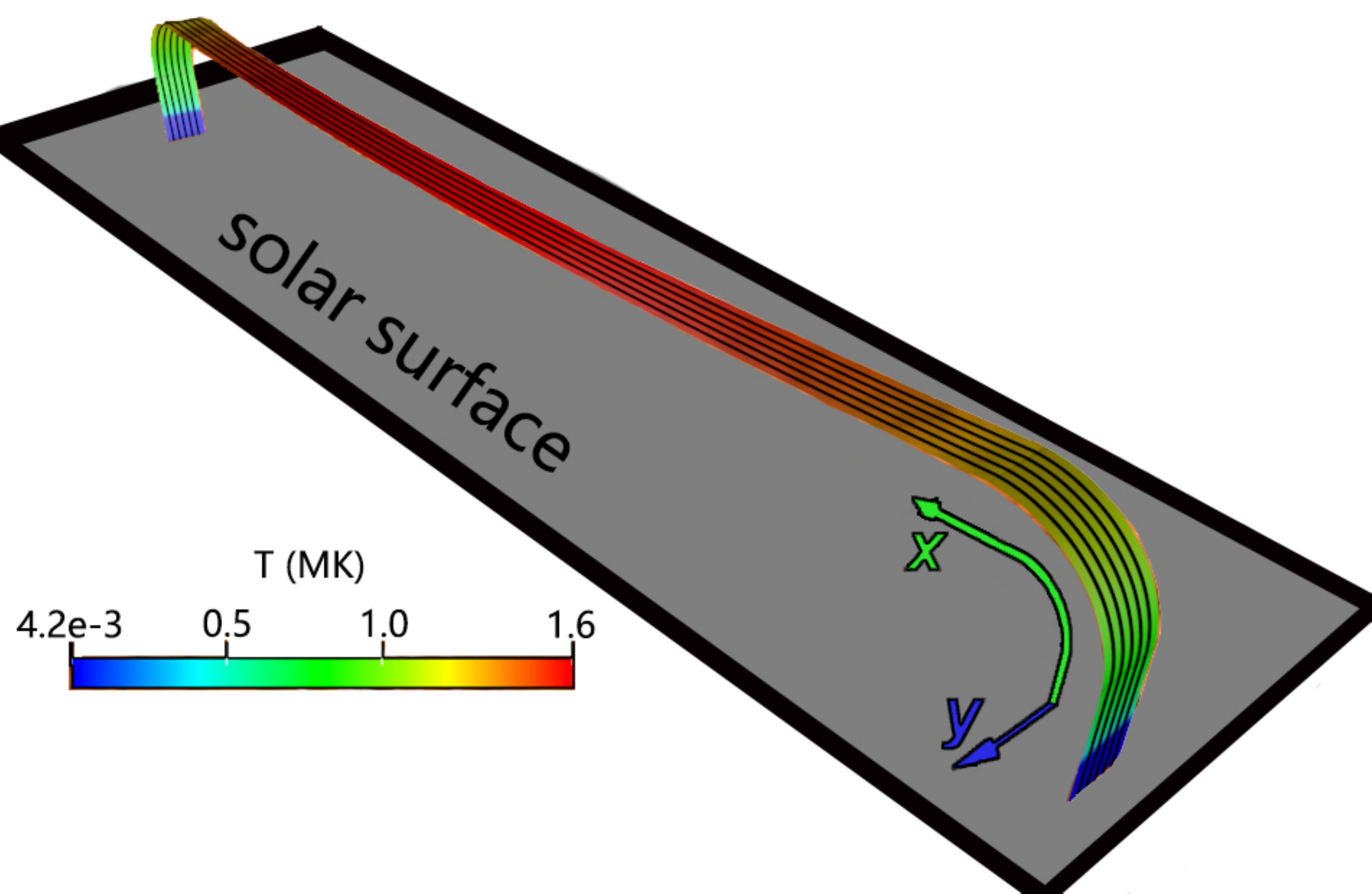}
\caption{\textbf{Magnetic configuration of our 2D model rendered in the 3D space.} The top part of the slab is quasi-parallel to the solar surface. The arrows at the lower-right corner show the $x-$ and $y-$directions. The arcade is colour-coded by the temperature in the initial condition, where the purple part is located in the solar chromosphere and the green/red parts are in the corona. Black lines indicate the orientation of the magnetic field lines.
} \label{fig1}
\end{figure}

\begin{figure}
\centering
      \includegraphics[width=0.60\textwidth]{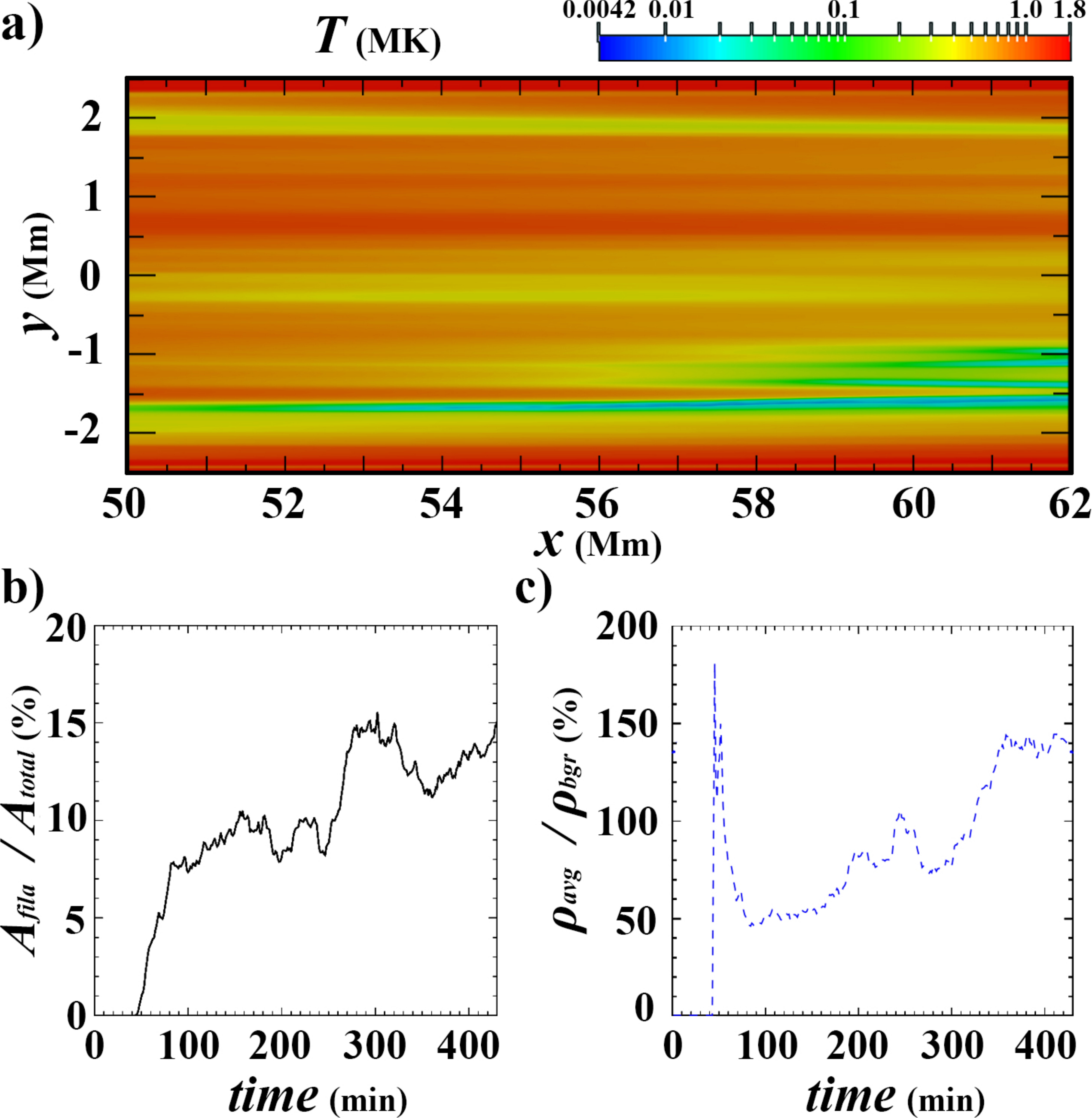}
\caption{\textbf{Evolution of the filament.} \textbf{a}, Temperature distribution at the centre of the simulation box at $t = 44$ min, when the first condensation happens. \textbf{b}, Time evolution of the relative area of the filament represented by the ratio between the area covered by filament threads ($A_{fila}$) and the total area of the corona ($A_{total}$). \textbf{c}, Time evolution of the averaged density of the filament ($\rho_{avg}$) compared with the background density ($\rho_{bgr}$), namely, $\rho_{avg} / \rho_{bgr}$.
} \label{fig2}
\end{figure}

\begin{figure}
\centering
      \includegraphics[width=0.60\textwidth]{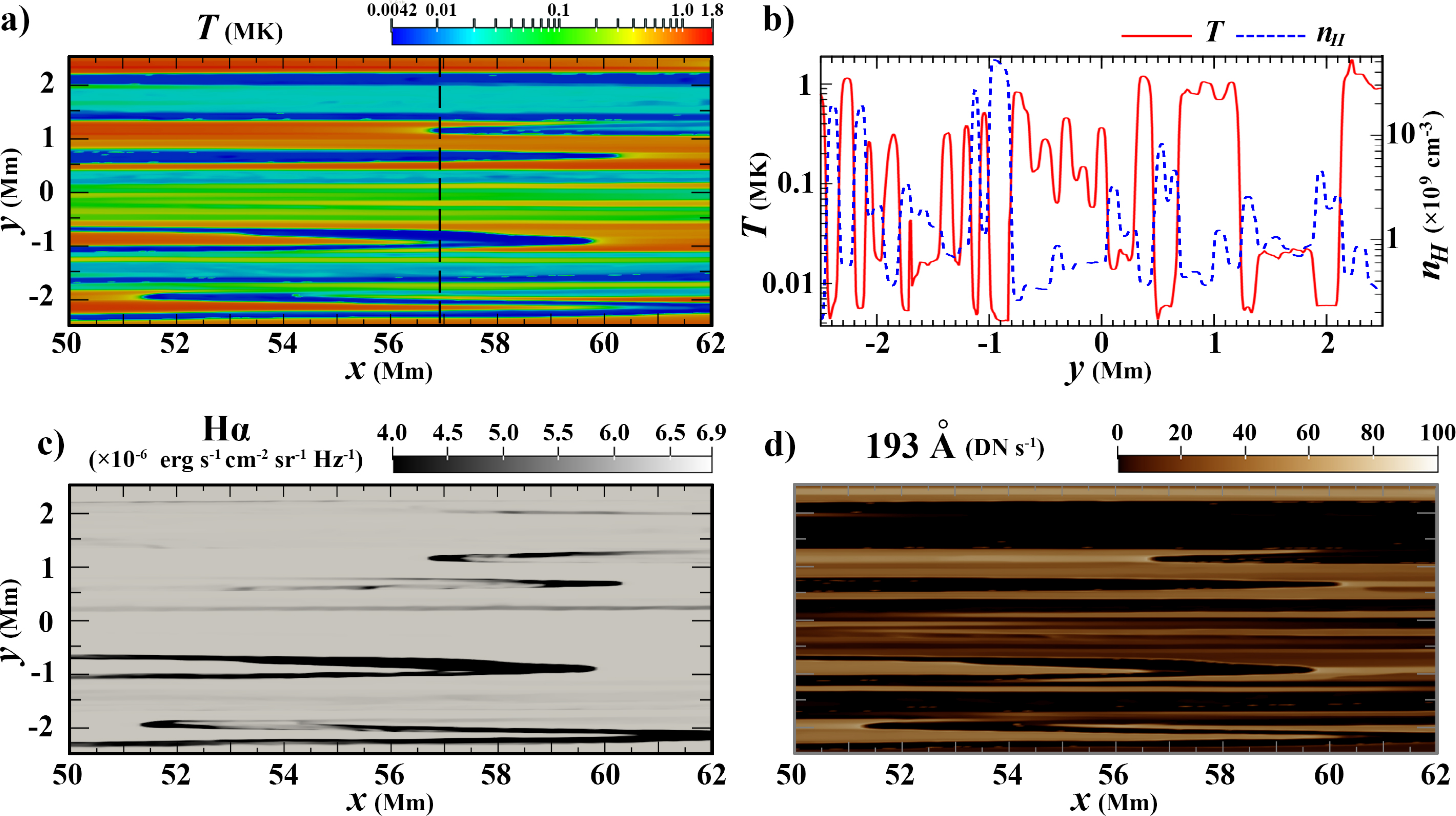}
\caption{\textbf{Distributions of various parameters in the solar corona at $t = 259$ min.} \textbf{a}, Temperature distribution near the centre of the simulation box. Temperature is shown in the log scale to highlight the threads. \textbf{b}, Temperature (red solid line) and number density (blue dashed line) distributions along the vertical slice cut in {\bf a}. The vertical coordinate is plotted in the log scale. \textbf{c}--\textbf{d}, Synthesized H$\alpha$ image and EUV 193 \AA~image corresponding to the area in {\bf a}.
} \label{fig3}
\end{figure}

\begin{figure}
\centering
      \includegraphics[width=0.60\textwidth]{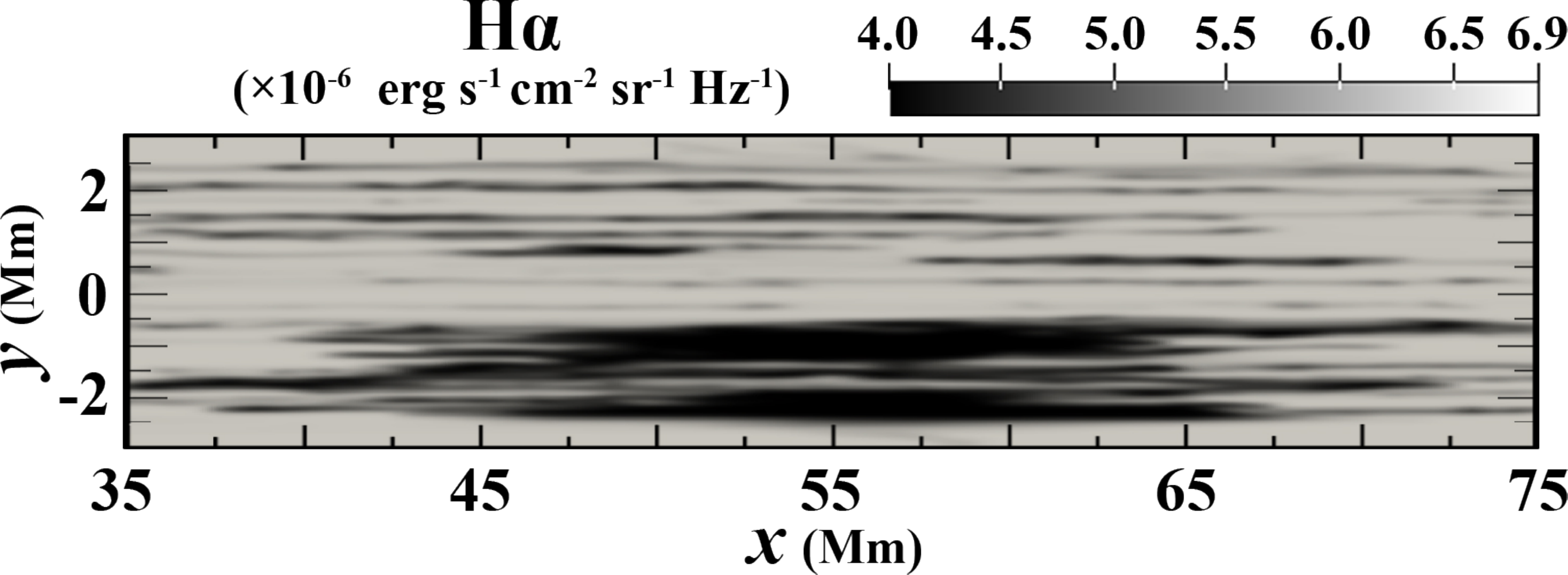}
\caption{\textbf{Synthesized H$\alpha$ image after superposing the results at different times.} The dark fibrils are filament threads.
} \label{fig4}
\end{figure}

\begin{figure}
\centering
      \includegraphics[width=0.60\textwidth]{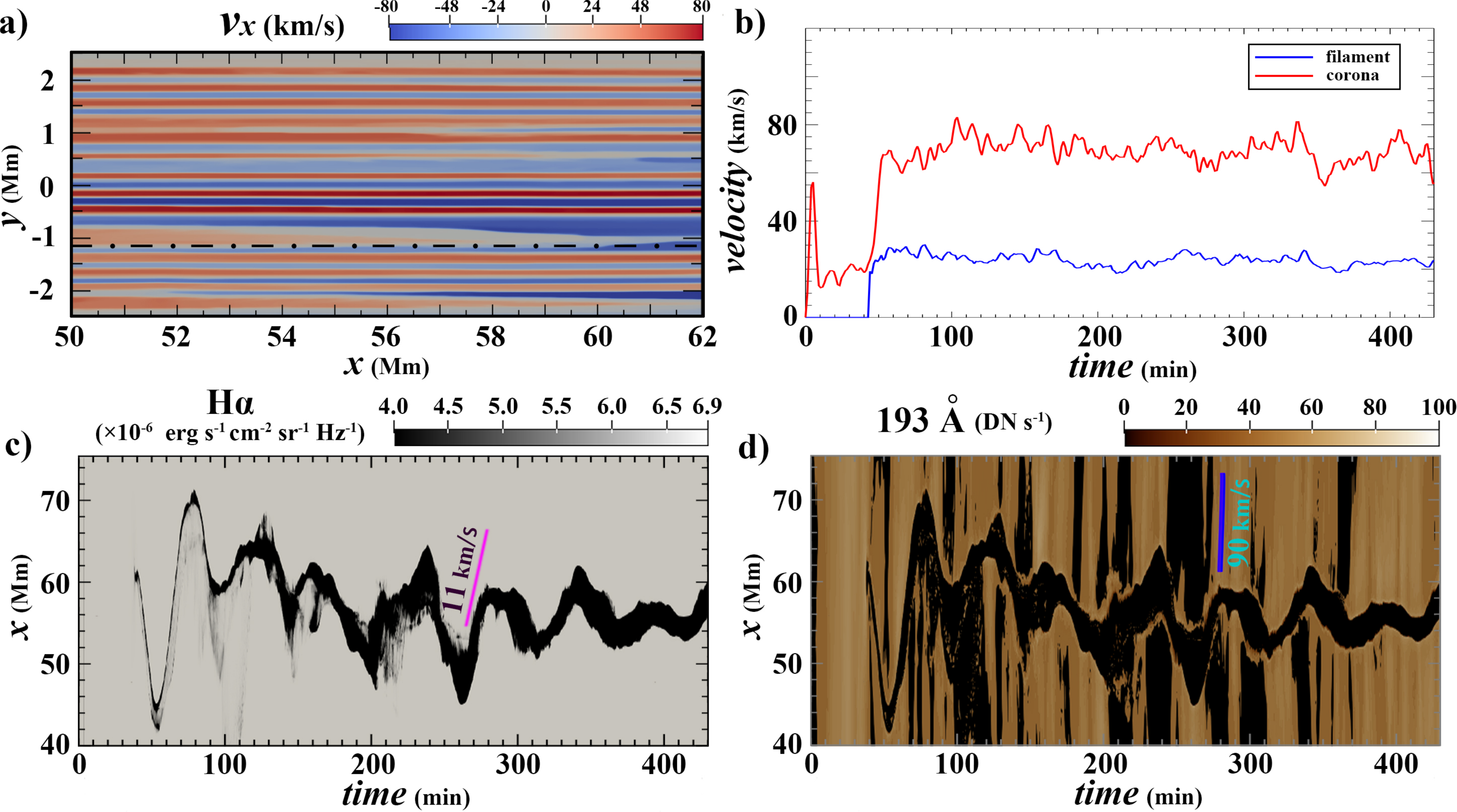}
\caption{\textbf{Dynamics of the filament threads and the ambient solar corona.} \textbf{a}, Velocity distribution within the same area as Fig. \ref{fig3}. \textbf{b}, Evolution of the averaged velocities inside the filament threads (blue line) and outside the filament threads (red line) for the materials moving in the positive $x$-direction. \textbf{c}--\textbf{d}, Evolution of the emission intensity along a slice cut in the H$\alpha$ and 193 \AA~synthesized images. The slice is marked by the horizontal dotted-dashed line in {\bf a}. The slopes of the magenta lines characterize the velocities of the H$\alpha$ thread and 193 \AA\ moving features, respectively.
} \label{fig5}
\end{figure}

\begin{figure}
\centering
      \includegraphics[width=0.60\textwidth]{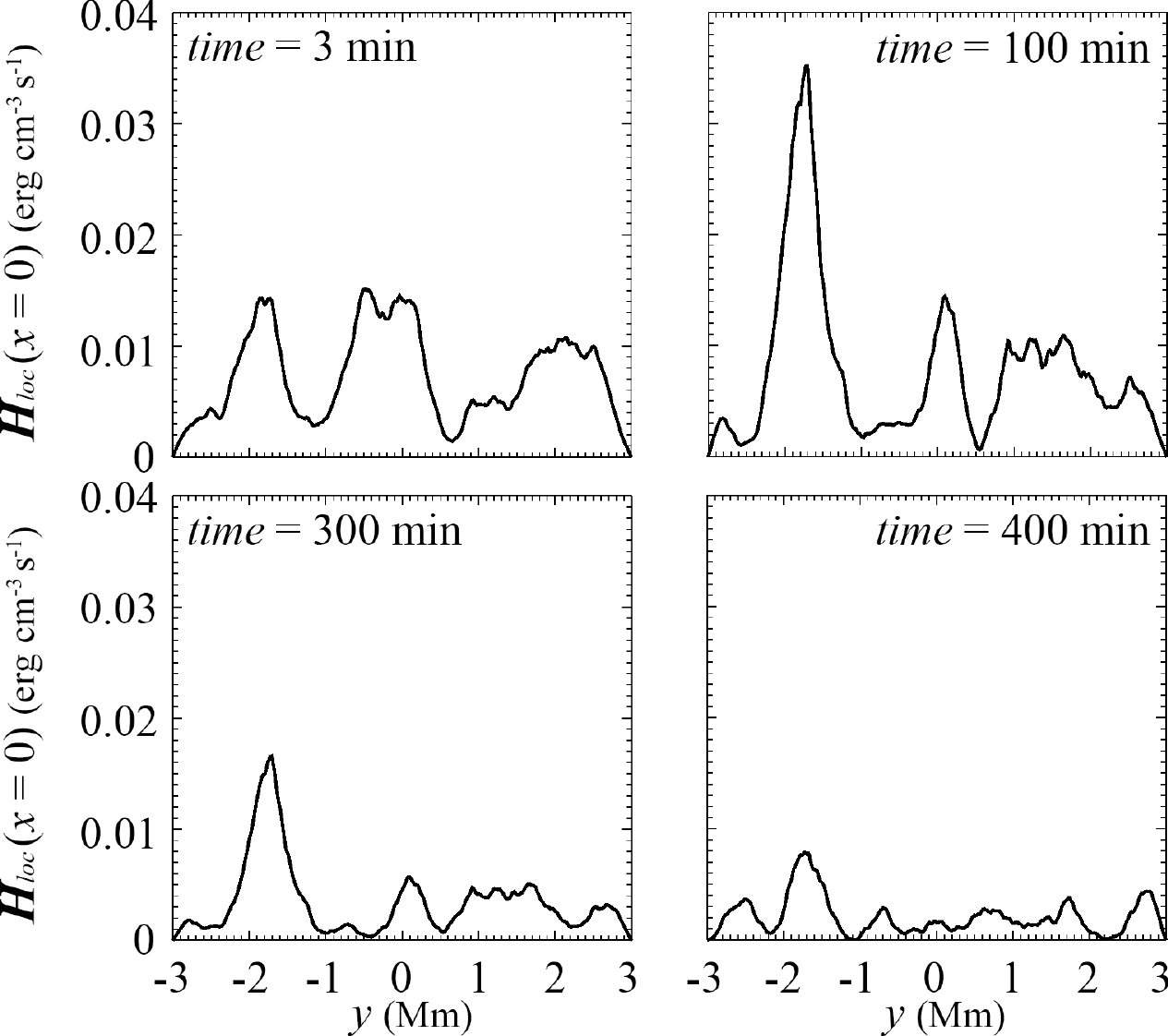}
\caption{\textbf{Four snapshots of the heating function distribution at the footpoints along $x=0$ in our simulation.}
} \label{fig6}
\end{figure}

\clearpage

\section*{Methods}

\textbf{Simulation setup.}
The magnetic configuration of a solar filament is usually considered to be either a sheared arcade or a flux rope \citep{vial2015, ouy2017}. To focus on the effect of randomised heating, here we choose to perform the simulation in a simple and straightforward geometry, namely, a sheared arcade. So that we can circumvent the difficulty raised by a complex geometry. For the flux rope--type magnetic configuration, the dynamics influenced by the field-aligned gravity would actually be the same.

In principle, studying the filament formation and evolution in a sheared arcade as illustrated in Fig. \ref{fig1} is a full 3D problem. However, with the current computational power, it is difficult to have high spatial resolution in order to resolve the fine structure of a filament in 3D MHD simulations, especially when radiation and field-aligned heat conduction are included. Therefore, we reduce the problem into a 2D one. To do so, we extend the widely used 1D flux tube model \citep{anti1991, karp2001, xia2011} to a 2D flux sheet model, as illustrated in Fig.~\ref{fig1}, where the magnetic field (black lines) is along the elongated direction of the flux sheet. The flux sheet is a curved surface, where the top part is quasi-parallel to the solar surface, and the two legs are anchored at the solar photosphere. The magnetic field lines are allowed to move in the sheet plane, but not in the direction perpendicular to the flux sheet. The size of the flux sheet is $0 \le x \le 112$ Mm and $-3 \le y \le 3$ Mm, which is typical in the solar filament channels. The two legs of the flux sheet, i.e., the parts near the two narrow edges, are perpendicular to the solar surface with a height of 8 Mm, and the top part of the flux sheet is nearly horizontal, containing a shallow dip with a depth of 3 Mm. For simplicity, the dipped magnetic field is described with a sine function as used in 1D flux tube models \citep{xia2011}. Although the existence of magnetic dips is unnecessary for filaments \citep{wang1999, karp2001}, they do exist in many events as implied by filament longitudinal oscillations \citep{chen2014}. The weakly dipped top part of the flux sheet is connected to the two leg parts with a quarter circle whose radius is 5 Mm. As a result, the central part of the magnetic dip is at an altitude of 10 Mm. Once the magnetic configuration is chosen, the gravity distribution in the flux sheet plane is determined. MHD equations are numerically solved in the 2D Cartesian coordinates with the resulting non-uniform gravity.

The initial magnetic field is assumed to be uniform with a strength of 5 G. To construct the atmospheric model, partial ionisation of hydrogen is considered, and the pressure gradient force balances the gravity in the hydrostatic equation. To compensate for the radiative cooling, a background heating is imposed in order to maintain the hot corona. With the prescribed quasi-equilibrium atmosphere, we perform numerical simulations for about 1 hour in order for the system to relax to a thermal equilibrium state along with force balance. After the relaxation process, the typical temperature of the coronal plasma is 1.6 MK, and the corresponding number density of the electrons is $3\times 10^8$ cm$^{-3}$. These lead to plasma $\beta$ being 65 at the footpoint ($x=0$) and $\sim$0.2 near the loop top.

We perform our simulations in the framework of the chromospheric evaporation--coronal condensation model. This model requires an extra localised heating at the footpoints of the magnetic field. This heating is usually thought to be 1--3 orders stronger than the background heating \citep{with1977, asch2001, xia2011}, which actually has a wide range, and it should decay rapidly along the magnetic field line \citep{xia2011}. Such a localised heating is usually imposed uniformly near the footpoints in previous 2D or 3D simulations \citep{xia2012, kepp2014, xia2016}. To apply our idea to the evaporation--condensation model, we specify the localised heating to be random along the $y-$axis, with a power-law spectrum in the spatial dimension. It is found that the final result is not sensitive to the choice of the spectral index. To make it more realistic, the heating is also random in time. The setup of this randomised heating can be found in the \textbf{Randomised heating} paragraph. Once the randomised heating is introduced near the footpoints of the magnetic field lines, we investigate the evolution of the magnetized plasma in the flux sheet.

The open source adaptive mesh refinement versatile advection code (MPI-AMRVAC 2.0)\citep{xia2018} is used to numerically solve the MHD equations. We use a uniform mesh, with $1,920\times 576$ grid points in the $x-$ and $y-$directions, respectively, so that the resolution is 58 km in the $x-$direction and 10 km in the $y-$direction. Since the width of threads is usually observed to be more than 100 km (which might be limited by the resolution of instruments), a resolution of 10 km is enough to resolve it. In this simulation, field-aligned gravity, optically thin radiation, and field-aligned thermal conduction are all considered.

\textbf{Ionisation.}
In our simulations, the plasma is partially ionised in order to synthesize H$\alpha$ images based on the simulation results. Thus, the number density of hydrogen at the first and second excitation levels ($n_2$ and $n_3$) is what we are really concerned about. We calculate the number density at different excitation levels by directly using the ionisation degree table $X = X(p, T)$ in Reference \citep{hein2015}. This table is specially designed for the filament environment at an altitude around 10,000 km, which is exactly our case.
\\

\textbf{Numerical Strategy.}
We use the second-order three-step Harten-Lax-van Leer scheme \citep{hart1983} with a third-order slope limiter \citep{cada2009} to solve the MHD equations.
Such a total variation diminishing (TVD) type scheme has the advantage in avoiding numerical overshoots when the physical quantities change abruptly in space. As compensation, it has the disadvantage of numerical diffusion which might oversmooth the numerical results. To deal with the optically thin radiative cooling, a usual approach is to make use of the table calculated by Reference \citep{colg2008}, which uses a complete and self-consistent atomic data of 15 coronal elements and their collisional data. This table works well in previous works \citep{xia2011, xia2012, xia2016}. However, this table may underestimate the cooling rate when temperature is below $10^5$ K. As a result, the formed filament usually has a temperature around $2 \times 10^4$ K \citep{anti2000}. In order to circumvent this issue, we choose to use the cooling rate described in the Reference \citep{schu2009} to deal with temperature below $10^5$ K until 4,170 K. This term is added by using an exact integration scheme \citep{town2009}. For the parallel thermal conduction, we use the classical Spitzer type coefficient ${\kappa _{{\rm{//}}}}$. This term is dealt by the Super Time-Stepping scheme as many other codes.
Perpendicular thermal conduction is not added explicitly. The numerical perpendicular thermal conduction $\kappa _ \bot$ is estimated to be of the order of $10 ^ {-7}{\kappa _{{\rm{//}}}}$ in our tests, therefore it can be neglected.\\

\textbf{Randomised heating.}
As mentioned in the first section of Methods, our heating term $H(x,y,t)$ is composed of two parts, the background heating $H_{bgr}$ and the localised heating $H_{loc}$. The background heating at one footpoint $x=0$ is $H_{bgr} = H_0\exp(-x/l_0)$, where $H_0=3 \times 10 ^ {-4}~\unit{erg}\unit{cm^{-3}}\unit{s^{-1}}$ and $l_0 = 56$ Mm \citep[refs.][]{with1977, with1988}, and its counterpart at the other footpoint is identical. For the localised heating $H_{loc}$, we separate the function into two parts, a time-varying part and a spatially-varying part, i.e., $H_{loc} = H_1H_x(x)H_y(y)H_t(t)$. $H_1$ is chosen to be $H_1=10 ^ {-2}~\unit{erg}\unit{cm^{-3}}\unit{s^{-1}}$ \citep[refs.][]{with1977, asch2001}, which is similar to previous works. First, we discuss the spatial distribution. Since we focus on the effect of randomised heating on the $y-$direction only, the heating profile in the $x-$direction $H_x$ is treated in the traditional way, namely, a similar form to the background heating, but with a much shorter scale height, which is set to be 2 Mm in our simulations.
For $H_y$, we decompose the heating into a sum of all wavelengths (from the simulation box down to the grid size), i.e., $\left( {\sum\limits_i {{A_i}\sin (2\pi y/{\lambda_i} + {\phi_i})} } \right)^2$, where $A_i$ is the amplitude and $\lambda_i$ is the wavelength of the heating profile, and $\phi_i$ the phase angle. For a white noise, $A_i$ should be a constant and $\phi_i$ is a random number. However, as we know, the stochastic nature of the motions and the large Reynolds numbers on the Sun suggest that the heating in the solar atmosphere has a power-law spectrum, which means ${A_i} \propto {\lambda ^s}$, where $s$ is the spectral index. We choose $s = 3/4$ in our simulations, which corresponds to the traditional magnetic turbulence spectrum. We also did many other tests with $s$ being changed in a wide range $0.2 \le s \le 2$, and it is found that the results do not change too much. Besides, to avoid the influence from the boundaries, $H_y$ decreases linearly to 0 near the two ends of the $y$-axis. Regarding the time-varying part of the heating function, we simply assume that it has an episodic Gaussian profile, which is widely used to fit the energetic events in the solar atmosphere. The duration of each episode of heating is chosen to be 5 minutes with a little variation. This is the typical time scale for lower atmosphere motions as well as their typical correlation time. With this assumption, the expression of the localised heating term is a sum of many Gaussian functions of time, i.e., $H_{loc} = \sum\limits_i{H_{i1}H_{it}(t)H_{ix}(x)H_{iy}(y)}$. Four snapshots of the time evolution of this localised heating at one footpoint is displayed in Fig. \ref{fig6}. \\

\textbf{Synthesized image.}
The 193 \AA\ EUV band is usually considered to be optically thin. However, the emission of this line could be reduced by the presence of solar filaments due to absorption and volume blocking \citep{anze2005}. In order to take this factor into consideration, we solve the radiative transfer equation. By assuming a uniform source function, we have the well-known solution ${I_{193}} = {S_{193}}(1 - {e^{{-\tau_{193}}}}) + {I_{bgr193}}{e^{{-\tau_{193}}}}$. Since the background emission ${I_{bgr193}}$ can be assumed to be any constant, the unknowns here are the source function $S_{193}$ and the optical depth $\tau_{193}$. The source function $S_{193}$ is the ratio of the emission coefficient $j_{193}$ and the absorption coefficient $\alpha_{193}$. $j_{193}$ could be approximated by the optically thin radiation $n_{\rm e}n_{\rm H}G(T)$, where $G(T)$ is the response function from the CHIANTI database \citep{dere19}, $n_e$ and $n_{\rm H}$ are the number density of electrons and hydrogen, respectively, and all emission lines near 193 \AA\ are considered. We calculate $\alpha_{193}$ from the absorbing cross section, which could be found in the literature \citep{anze2005}. It is noted that the 193 \AA\ absorption is mainly caused by the photoionisation of neutral hydrogen plus neutral helium. Since only hydrogen is considered in our simulation, the accurate $\tau_{193}$ corrected for helium contribution is expressed as the formula $\tau_{193}=0.1n_{\rm H} L_0 \sigma_{\rm {He}}+n_{\rm H} L_0 i \sigma_{\rm H}$, where 0.1 is the abundance of helium, the multi-layer threads are assumed to have a thickness of $L_0$=2.5 Mm, $i$ is the ionisation degree of hydrogen, and $\sigma_{\rm H}$ and $\sigma_{\rm {He}}$ are the photoionisation cross sections for hydrogen and neutral helium, respectively. The values of $\sigma_{\rm H}$ and $\sigma_{\rm {He}}$ and the details of the radiative transfer calculation can be found in Ref. \citep{anze2005}.

The H$\alpha$ line at 6562.8 \AA\ is an optically thick line. That means we should have a whole 3D data cube and solve the radiative transfer equation. Assuming a uniform source function, the well-known solution $I_{H\alpha} = S(1-e^{-\tau}) + I_{bgr}e^{-\tau}$ could be obtained, where $S$ is the source function and $\tau$ is the optical depth. Suppose that the single layer in our simulations has a thickness of 2.5 Mm, and neglect the H$\alpha$ emission from the corona, a fast approximate radiative transfer method for visualizing filaments \citep{hein2015} is used to calculate the optical depth $\tau$. The table for filaments at an altitude of 10 Mm is used so that the results are consistent with the ionisation degree table used above. \\

\textbf{Data availability.} Any snapshots of simulation data (2 TB in size) are available on request to Y. H. Zhou (yuhaozhou1991@gmail.com). The data for the figures are avaliable at \url{ https://doi.org/10.6084/m9.figshare.11956482}.

\textbf{Code availability.} The code we use is the open-source code AMRVAC 2.0, which can be downloaded directly from its website \citep{xia2018}.

\end{document}